\documentclass{optica-article}

\journal{opticajournal} 

\articletype{Research Article}

\usepackage{xspace}
\usepackage{amsmath}
\usepackage{graphicx}
\usepackage{xcolor}
\usepackage{comment}
\usepackage{multirow}
\usepackage{rotating}
\usepackage{diagbox}
\usepackage{booktabs}

\newcommand{\invitro}{\textit{in vitro}\xspace}

\newcommand{\enface}{\textit{en face}\xspace}
\newcommand{\um}{\(\muup\)m\xspace}
\newcommand{\uM}{\(\muup\)M\xspace}
\newcommand{\us}{\(\muup\)s\xspace}
\newcommand{\invum}{\(\muup\)m$^{-1}$\xspace}
\newcommand{\OCDS}{OCDS\textsubscript{\textit{l}}\xspace}

\graphicspath{{./figure}{.}}

\begin{document}

\title{Dynamic full-field swept-source optical coherence microscope for cellular-resolution, long-depth, and intratissue-activity imaging}

\author{Nobuhisa Tateno,\authormark{1} 
		Yue Zhu,\authormark{1}
		Suzuyo Komeda,\authormark{1}
		Mahiro Ishikawa,\authormark{1}
		Xibo Wang,\authormark{1} 
		Ibrahim Abd El-Sadek,\authormark{1,2} 
		Rion Morishita,\authormark{1} 
		Atsuko Furukawa,\authormark{3} 
		Satoshi Matsusaka,\authormark{3}
		Shuichi Makita,\authormark{1,*} 
		and Yoshiaki Yasuno\authormark{1,**}}

\address{\authormark{1}Computational Optics Group, University of Tsukuba, Tsukuba, Ibaraki, Japan.\\
\authormark{2}Department of Physics, Faculty of Science, Damietta University, New Damietta City, Damietta, Egypt.\\
		\authormark{3}Clinical Research and Regional Innovation, Faculty of Medicine, University of Tsukuba, Tsukuba, Ibaraki, Japan.
}

\email{\authormark{*}shuichi.makita@cog-labs.org}
\email{\authormark{**}yoshiaki.yasuno@cog-labs.org}

\begin{abstract*} 
	Optical coherence tomography (OCT) microscope (OCM) uses a high-numerical-aperture objective to achieve cellular-level lateral resolution.
	However, its practical imaging depth range is limited by the depth of focus (DOF).
	Although computational refocusing can potentially provide sharp images outside the DOF, signal reduction by the confocal effect still limits the imaging depth in practice in point-scanning OCT\@.
	In addition, standard OCT cannot visualize intra-tissue activities.
	To overcome these limitations, we demonstrated a spatially coherent full-field OCM (SC-FFOCM) with computational refocusing.
	In addition, a repetitive acquisition protocol was designed to visualize intra-tissue activities (i.e., dynamic OCT imaging).
	The in-focus lateral resolution is 1.4 \um, and the axial resolution is 6.5 \um (in air) at full-width at half-maximum intensity.
	Three-dimensional structure and the dynamic OCT imaging using SC-FFOCM with computational refocusing was applied to human breast adenocarcinoma spheroids (MCF-7 cell line).
	Volumetric dynamic imaging with cellular-level lateral resolution was demonstrated over the full depth of the spheroid.
\end{abstract*}

\section{Introduction}
In recent years, there have been several applications for the assessment of \invitro thick samples such as spheroids\cite{zoetemelkShortterm3DCulture2019} and organoids\cite{takebeVascularizedFunctionalHuman2013a}.
Non-invasive imaging methods are needed to track the time course of local changes in tissue, such as tracking the response of a local site to drug administration.
When imaging thick samples, it is essential to achieve cellular-level resolution and long imaging depths simultaneously.

One of the promising methods is optical coherence tomography (OCT)\cite{Huang1991}, which is a non-invasive and label-free imaging modality with a millimeter-scale imaging depth.
In general, OCT uses a low numerical aperture (NA) objective and a resulting weakly focused probe beam to achieve a long imaging depth.
Here, the long depth of focus (DOF) preserves the lateral resolution and signal magnitude along the deep imaging depth in a single B-scan image, however, the lateral resolution becomes low.
On the other hand, optical coherence microscopy (OCM)\cite{aguirreOpticalCoherenceMicroscopy2015} uses a high NA objective to achieve cellular-level lateral resolution, however such a configuration results in a short DOF\@.
Namely, the high lateral resolution and the long imaging depth cannot be obtained simultaneously\cite{drexler_vivo_1999, huber_three-dimensional_2005, rolland_gabor-based_2010, srinivasan_optical_2012}.
Therefore, high-resolution OCM is not well suited for volumetric imaging of thick samples.
This trade-off has been partially resolved by several computational refocusing methods\cite{liuComputationalOpticalCoherence2017a} such as the digital holographic method that inverts defocusing\cite{yasuno_non-iterative_2006,yu_improved_2007} and interferometric synthetic aperture microscopy (ISAM)\cite{ralston_inverse_2006,ralstonInterferometricSyntheticAperture2007}.  The depth range preserving high lateral resolution is extended.

While computational refocusing restores the lateral resolution by correcting defocus, there are still several problems in the case of widely used point-scanning OCT\@.
One problem is imperfect defocus correction.
According to a recent theoretical investigation, computational refocusing during point-scanning OCT cannot fully correct defocus for a high NA objective\cite{fukutake_four-dimensional_2025-1,fukutake_unified_2025-1,makita_image_2025-2}.
In addition, because the wavefront errors in the illumination and collection paths can interact, the relationship between wavefront aberrations and the phase error of the OCT signal is complicated \cite{southWavefrontMeasurementUsing2018,makita_image_2025-2}.
Another problem is optical loss.
The confocal pinhole (i.e., a single-mode fiber tip) reduces the total energy of the singly-scattered light from a scatterer as the defocus increases\cite{zhu2025theoreticalanalysisperformancelimitation}.
This results in a low signal-to-noise ratio (SNR) at the defocused region, even if the lateral resolution is restored by computational refocusing.

To overcome the above issues for \invitro thick tissue imaging, we applied spatially coherent full-field OCM (SC-FFOCM) with a high NA that is incorporated with computational refocusing.
Instead of scanning the focused beam in point-scanning OCT, a spatially coherent light source was used for flood-illumination of the sample.
This type of OCT is usually accomplished by using a wavelength-swept light source and a high-speed camera, and the sample is illuminated by a plane wave\cite{boninVivoFourierdomainFullfield2010,hillmannOffaxisReferenceBeam2017a}.
Because there is no confocal gating, the energy of the collected scattered light is not reduced by defocus.
In addition, since the illumination is a plane wave, the computational refocusing and aberration correction are simpler than those for point-scanning OCT\@.

Even if these problems are resolved by SC-FFOCM, there is an additional issue that standard OCT is only sensitive to sample structures.
Therefore, OCT cannot visualize intra-tissue activities.
Dynamic OCT (DOCT) is a combination of time-sequential OCT measurements and subsequent analyses of time-sequential signals and is used for visualizing intra-tissue activities\cite{Elsadek2020BOE,ElSadek2021BOE,leungImagingIntracellularMotion2020a,munterDynamicContrastScanning2020a,Morishita2023BOE}.

In this paper, we introduce an SC-FFOCM system with computational refocusing and DOCT imaging.
To achieve DOCT imaging, we designed a repetitive volume acquisition protocol suitable for SC-FFOCM\@.
SC-FFOCM combined with computational signal processing and DOCT imaging can visualize intra-tissue activity with cellular-level resolution.

\section{Principle and methods}
\label{sec:principle}
\subsection{System design}
\label{sec:systemDesign}
\begin{figure}
	\centering\includegraphics[width=13cm]{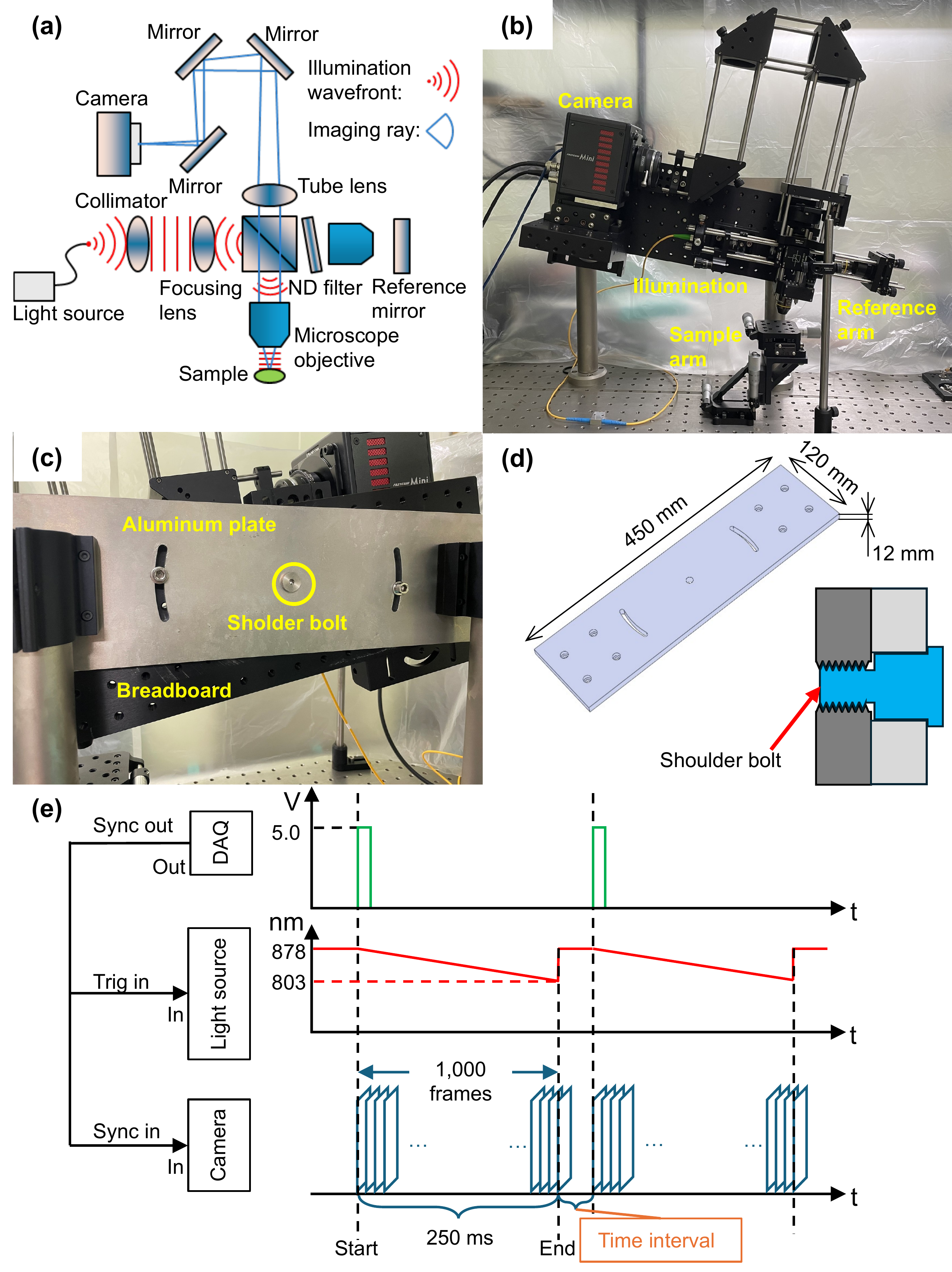}
	\caption{(a) Schematic of the optical design.
		(b) Front and (c) back appearances of the spatially coherent full-field optical coherence microscopy (SC-FFOCM) system.
		(d) Schematic of the custom aluminum plate and shoulder bolt.
		(e) Schematic of synchronization of light-source wavelength sweeping and data acquisition by the camera.}
	\label{fig:System}
\end{figure}
Figures \ref{fig:System}(a) and \ref{fig:System}(b) show the optical and mechanical designs of the custom-built SC-FFOCM system.
When imaging cultured samples, surface reflections from the culture medium sometimes go to the camera and consume the detection range of the camera.
To avoid these reflections, it is necessary to tilt the system without losing optical alignment.
Figures \ref{fig:System}(c) and \ref{fig:System}(d) show the back of the system and the mechanical components used to tilt the optical system.
For tilting, we combined a custom aluminum plate (MISUMI, Japan) and a shoulder bolt (DBS6-10-12; MISUMI, Japan).
The plate is $120 \times 450$ mm$^2$ with a thickness of 12 mm.
An optical breadboard (120 mm $\times$ 450 mm $\times$ 12 mm, MB1545/M, Thorlabs) is fixed to the aluminum plate with the shoulder bolt, and it allows the whole optics to be tilted up to $\pm$ 28 degrees.
In the specific configuration of this study, the system is tilted by 8 degrees.
To avoid vibrations from the fans of the light source and the computer, they are not placed on the same optical table as the OCM optics.
In addition, the camera fan is turned off during the imaging.

Figure \ref{fig:System}(e) shows the synchronization chart of light-source wavelength sweeping and data acquisition by the camera.
Both the light source and the camera are waiting for a trigger to start the wavelength sweep and acquisition, respectively.
A function generator (DAQ USB-6361-BNC, National Instruments, TX) generates and feeds a trigger to them for synchronization.
By assertion of the trigger, the light source starts one wavelength sweep with a sweeping speed of 300 nm/s, and the camera captures 1,000 images at a frame rate of 4,000 frames/s.
Triggers are applied 32 times for a dynamic OCT measurement, so that 32 OCT volumes are acquired in a sequence.
A time interval (i.e., a re-arm time) of 0.5 ms is required between the end of the light source sweep and the next trigger pulse.
The captured interferometric images saved in the camera are then transferred to the PC via a Gigabit Ethernet connection.

The sweeping light source (BS-840-1-HP, Superlum, Ireland) sweeps the wavelength with the range of 75 nm centered at 840 nm.
The light beam is collimated by a collimator lens ($f$ = 50 mm) and then made convergent by a focusing lens ($f$ = 100 mm).
The converged beam is split into a sample arm and a reference arm by a beam splitter, and then re-collimated by a microscope objective (NA = 0.3, effective focal length = 18 mm, RMS10X-PF, Olympus, Japan) to illuminate the sample.
The beam diameter on the sample is 2.34 mm, with a probe power of 5.23 mW.
The optical power of the reference beam is adjusted by a variable neutral density (ND) filter (optical density (OD) = 1.0, NDL-10S-2, Thorlabs).
The in-focus lateral resolution at the full-width at half-maximum (FWHM) intensity is 1.4 \um, and the axial resolution is 6.5 \um (FWHM intensity in air).
The DOF is 16.0 \um.

The sample is imaged with a high-speed two-dimensional (2D) CMOS camera (maximum frame rate = 6,400 frames/s for 1024 $\times$ 1024 pixels, FASTCAM Mini AX100, Photron, Japan) by the microscope objective and a tube lens ($f$ = 750 mm, 2-in diameter, ACT508-750-B-ML, Thorlabs); the magnification is 42.
The camera pixel size is 20 \um $\times$ 20 \um, and the sensor resolution is 1024 $\times$ 1024 pixels.

The maximum lateral field of view (FOV) is 0.49 mm $\times$ 0.49 mm (1024 $\times$ 1024 pixels) for the static (not dynamic) OCT measurement, while it is reduced to 0.43 mm $\times$ 0.37 mm (896 $\times$ 768 pixels) for DOCT imaging due to the limited on-camera memory.
The details of the DOCT-imaging mode are described in Section \ref{sec:principleDoct}.
According to the Nyquist theorem, computational refocusing can be successful if the cutoff spatial frequency limited by the NA of the objective is smaller than half of the spatial sampling frequency determined by the pixel separation on the sample \cite{zhu2025theoreticalanalysisperformancelimitation}.
The present SC-FFOCM setup satisfies this requirement.
For both OCT and DOCT imaging, the spectral frames are acquired at 4000 frames/s.

\subsection{Aberrations and phase error}
\label{sec:Aberration}
Computational refocusing with SC-FFOCM has an inherent advantage over point-scanning OCT\@.
Namely, the phase error in the spatial frequency domain due to the defocus was not affected by wavefront aberrations of the illumination optics.

To explain it theoretically, we first explain the pupil and aperture concepts to describe the relationship between aberrations and phase errors in the OCT image spectrum.
Each of the illumination and collection optics has its own pupil function (i.e., illumination and collection pupils) (see, for example, Refs.\@ \cite{zhu2025theoreticalanalysisperformancelimitation, fukutake_unified_2025-1,fukutake_four-dimensional_2025-1}), and the pupil functions represent the transfer of information via respective optics in the spatial frequency domain.
Wavefront aberrations of each optics can be represented as the phase errors in the pupil functions.

An aperture function is a function in the spatial frequency domain and is defined as the convolution of the illumination and collection pupil functions.
And the spatial frequency spectrum of the complex OCT image is the product of the aperture function and the spatial frequency spectrum of the sample structure.
Hence, the aperture function limits and modulates the information of the sample in the image.
Specifically, the phase of the aperture function becomes the phase error of the spatial frequency spectrum of the complex OCT image.

\begin{figure}
	\centering\includegraphics[width=13cm]{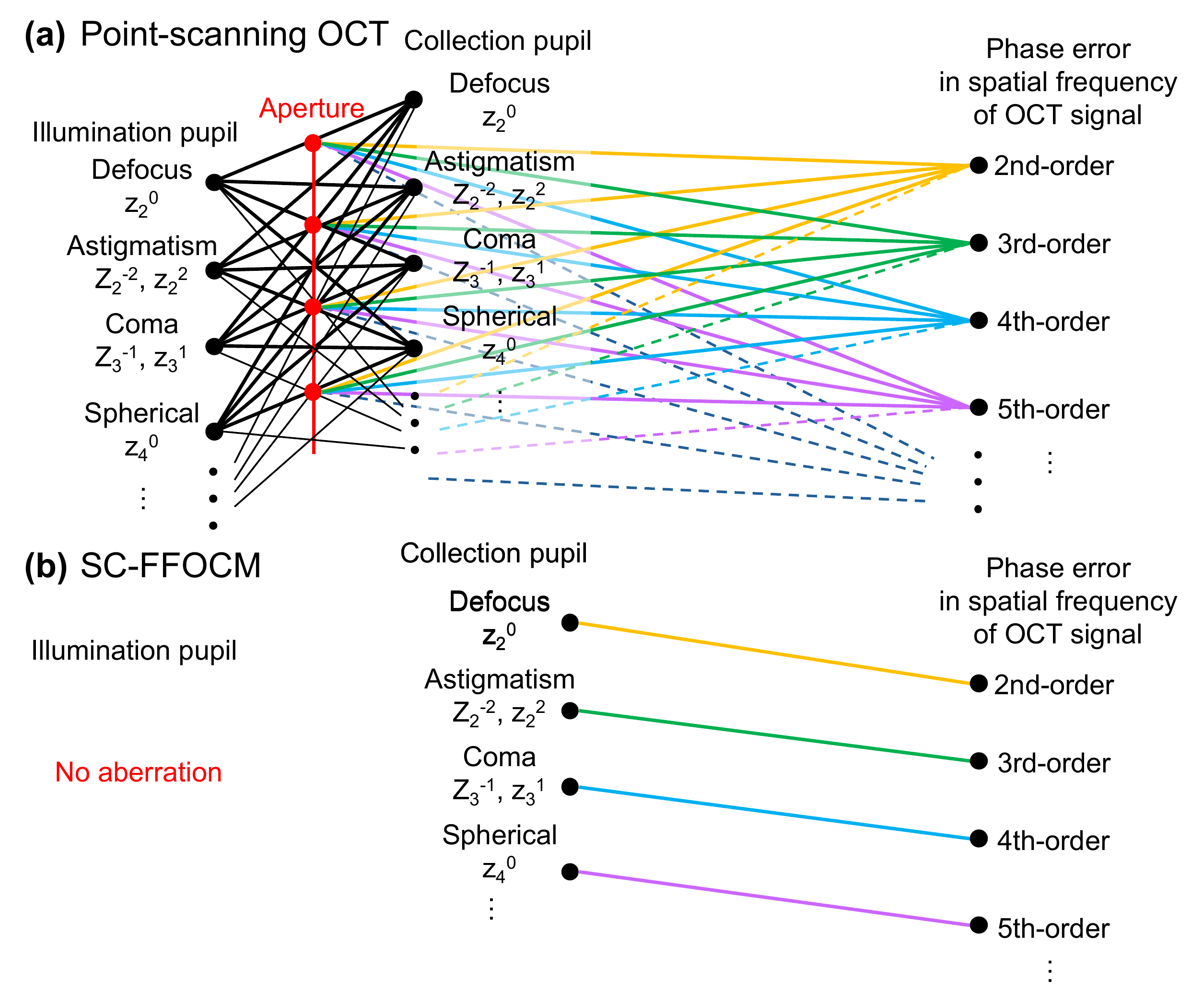}
	\caption{Schematic of the relationship between aberration and phase error in the spatial frequency spectrum of the OCT signal in (a) point-scanning OCT and (b) SC-FFOCM.}
	\label{fig:Aberration}
\end{figure}

In general, the illumination and collection optics share the same optics in point-scanning OCT\@.
Because the illumination and collection pupil functions are convolved, the respective aberrations interact with each other \cite{southWavefrontMeasurementUsing2018,makita_image_2025-2}.
This process causes a complicated phase pattern of the aperture function, and hence a complicated phase-error pattern of the spatial-frequency spectrum of the image.
It is noteworthy that the order of the aberration no longer corresponds to the order of the phase error as schematically depicted in Fig.\@ \ref{fig:Aberration}(a).
Namely, low-order aberration also partially corresponds to high-order phase-errors as it interacts with high-order aberrations.

On the other hand, the SC-FFOCM illuminates the sample with a plane wave, and hence its illumination pupil function is a delta function.
Here, the wavefront error on the illumination pupil function becomes only a phase offset and is negligible, and hence, the aperture function becomes identical to the collection pupil function except for the ineffectual phase offset.
In this case, the wavefront aberrations of the collection pupil function directly correspond to the phase error of the spatial-frequency spectrum of the image as depicted in Fig.\@ \ref{fig:Aberration}(b).
Namely, each order of the phase error directly corresponds to each order of the aberrations.
Specifically, in SC-FFOCM, the defocus affects only the second-order phase error.

\subsection{Decay of detected scattered light energy}
\label{sec:confocality}
Because the single-mode fiber tip is practically a confocal pinhole in point-scanning OCT, the total detectable energy of the single-scattering light decreases as the defocus amount increases as \cite{buist_theoretical_2024,zhu2025theoreticalanalysisperformancelimitation}
\begin{equation}
	I_\mathrm{PS}\ (z_d\ ) \propto \left[1 + \left(\frac{z_d}{z_R}\ \right)^2\right]^{-1},
\end{equation}
where $z_d$ is the defocus amount, and $z_R$ is the Rayleigh length.

For example, in point-scanning OCT, the total energy decreases by approximately 40 dB for a defocus amount of $20 z_R$.
On the other hand, because the SC-FFOCM does not have a confocal pinhole, all the scattered light can be captured and contribute to the interferometric signal formation.

In SC-FFOCM, the peak intensity of the OCT signal is still reduced by the defocus because of the blurring of the signal.
However, since the total light energy is preserved as we discussed above, the computational refocusing (Section \ref{sec:Refocusing}) can retrieve the original peak intensity.

\subsection{Signal processing for OCT and DOCT reconstruction}
\subsubsection{Raw-OCT reconstruction}
The obtained spectrum is linearly rescaled with respect to $k$ before the Fourier transform.
Here, the nonlinearity of the wavelength scan is calibrated by a fringe-analysis based calibration method \cite{yasunoThreedimensionalHighspeedSweptsource2005}, and the original 1,000 interference frames were resampled into 1,024 $k$-linear frames.
The raw OCT volumes are then obtained by fast Fourier transform (FFT) along $k$.

\begin{figure}
	\centering\includegraphics[width=10cm]{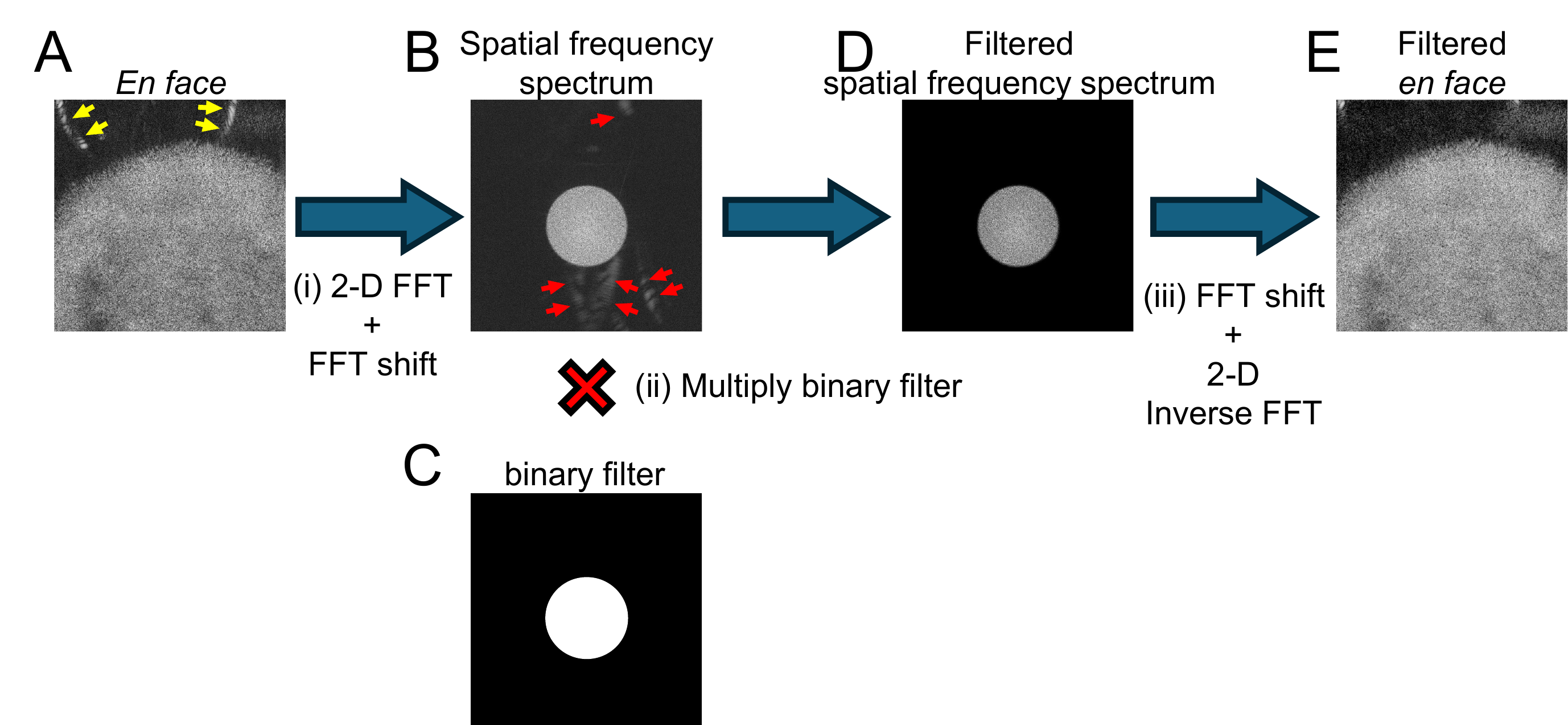}
	\caption{Schematic of artifact removal flow using a binary filter.
		The signals indicated by the red arrows are assumed to be artifacts in the high-frequency domain outside of the cutoff frequency.
	}
	\label{fig:Artifact}
\end{figure}
In SC-FFOCM, stray light, such as reflections from system components, is not filtered out by the confocal effects and hence causes artifacts in OCT images.
Here we removed these artifacts by a numerical binary spatial-frequency filtering as depicted in Figure~\ref{fig:Artifact}.
Here, the yellow arrows in [Fig.\@ \ref{fig:Artifact}(a)] indicate the artifacts, and the red arrows in [Fig.\@ \ref{fig:Artifact}(b)] indicate their corresponding spectral component.

The diameter of the spatial-frequency filter [Fig.\@ \ref{fig:Artifact}(c)] is selected based on the cut-off frequency of the objective, which is 0.357 \invum.
In the present implementation, the filter radius was selected to be 0.884 \invum, which is slightly larger than the cutoff frequency.
Note that the center of the binary filter is shifted horizontally by 0.100 \invum and vertically by -0.022 \invum to account for the shift of the spatial-frequency spectrum of the OCT image due to slight inclinations of the probe and the reference beams.

\subsubsection{Computational refocusing}
\label{sec:Refocusing}
The computational refocusing was performed by applying a phase-only spatial frequency filter \cite{kumarNumericalFocusingMethods2014, nakamura_complex_2007} in the spatial-frequency domain.
The phase-only spatial frequency filter is given by \cite{makita_image_2025-2}
\begin{equation}\label{eq:2}
	H^{-1}\left(f_x,\ f_y;\ z_d\ \right) = \exp\left[-i\pi\lambda_c\ \frac{z_d}{n} \left(f_x^2\ +\ f_y^2\ \right)\right]\ ,
\end{equation}
where $z_d$ is the defocus distance, $n$ is the refractive index of the sample, $f_x$ and $f_y$ are the spatial frequencies corresponding to the lateral positions $x$ and $y$, and $\lambda_c$ is the center wavelength of the probe beam.
By assuming that the depth position of the \enface plane is $z$ and expressing the OCT signal as $S(x, y, z)$, the refocused complex OCT signal $S'(x, y, z)$ is written as:
\begin{equation}
	S'(x,\ y,\ z\ )=
	\mathcal{F}^{-1} [\mathcal{F} [S(x,\ y,\ z\ )]\ H^{-1}\ (f_x,\ f_y;\ z_d\ )]\ ,
\end{equation}
where $\mathcal{F} [\quad]$ and $\mathcal{F}^{-1} [\quad]$ denote transversal 2D Fourier and inverse-Fourier transforms, respectively.
Prior to the computational refocusing, the \enface image field was extended using zero-padding to avoid aliasing artifacts at the periphery of the \enface field\cite{zhuComputationalRefocusingJones2022,makita_image_2025-2}.
In our particular implementation, numerical zero fields with 100-pixel width (48 \um in real space) were added to all four sides of the \enface field, and these zero fields are removed again after the refocusing operation.

Since the exact $z_d$, the focus position in the image, and the $n$ are not known, we model the $z_d/n$ as a linear function $a+bl$ where $l$ is the optical path length (OPL, i.e., the depth in the image), $a$ and $b$ are parameters.
In the refocusing process, the optimal $a$ and $b$ are selected as they sharpen eight \enface slices equally spaced in depth within the sample.

The estimation of the defocus parameter was performed as a two-step estimation of iterative image sharpness optimization to suppress the estimation time cost.
Here, the entropy-like image metric\cite{flores_robust_1992} of the maximum-intensity projection of the three consecutive linear intensity \enface OCT images is used to quantify the image sharpness, and the mean of the improvements of the entropy-like metrics at the eight depths is used as the optimization metric, the same as in Ref.\@ \cite{makita_image_2025-2}.
The metric was computed by using only the center region of the \enface images to avoid the influence of the field borders \cite{makita_image_2025-2}.
Namely, the peripheral region with a 60-pixel width (29 \um in the real space) was not used for the metric computation.

In the first step, the initial estimate of $a$ is obtained by setting $b = 0$ (i.e., the defocus is depth-independent).
The estimation was performed by brute-force searching with a searching range of $\pm 200\Delta l$, where $\Delta l$ is the OPL per single pixel.

In the second step, both $a$ and $b$ are simultaneously estimated to maximize the optimization metric.
Here, the estimate of $a$ obtained in the first step was used as the initial value of the iterative estimation (i.e., optimization).
In this process, the origin of $l$ is defined as the mean depth of the eight depths of the eight \enface slices.
The optimization was performed by the brute force searching followed by the Nelder-Mead simplex optimization.
The searching ranges of the brute force method were $\pm 40\Delta l$ around the initial value for $a$, and $\pm 1.2$ for $b$.
The searching range of $b$ was selected based on the theoretical prediction, $|b| \sim 0.55$ \cite{makita_image_2025-2}.
The estimation process including the first and second steps was implemented with the optimization modules in the SciPy Python library.

Finally, the \enface OCT images at all depths are refocused using the spatial-frequency filter [Eq.\@ (\ref{eq:2})] by substituting the estimated $a$ and $b$ into $(z_d/n) \equiv a+bl$.
It is noteworthy that the phase is perfectly stable over the lateral field, and hence the phase-error correction, which is necessary for point scanning OCT-based computational refocusing, was not applied.

\subsection{DOCT imaging}
\label{sec:principleDoct}
DOCT imaging was performed with a repetitive acquisition protocol designed for the SC-FFOCM\@.
In our particular implementation, 32 volumes are acquired sequentially.
The total acquisition time for the 32 volumes is 8.02 s and the volume repetition time is 250.5 ms.
This acquisition protocol is similar to the volumetric DOCT protocol of point-scanning OCT for spheroid and organoid \cite{ElSadek2021BOE, Morishita2023BOE}.

\begin{figure}
	\centering\includegraphics[width=7cm]{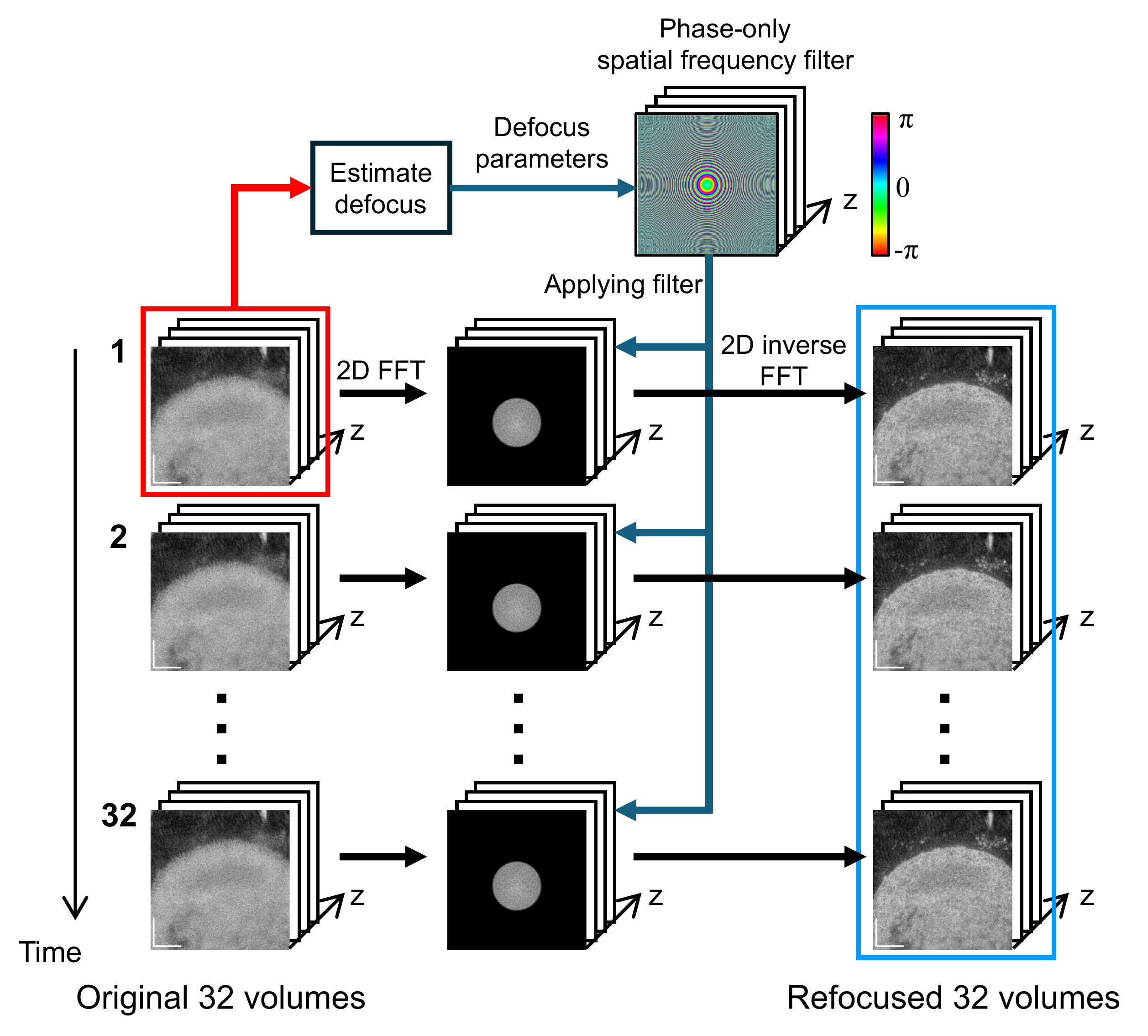}
	\caption{Schematic of multi-volume refocusing.
		The refocusing parameters are estimated by using the first OCT volume.
		The same filter was applied to all 32 volumes.
}
	\label{fig:Multi-volumeRefocusing}
\end{figure}
Before computing DOCT images, all 32 volumes are refocused as schematically explained in Fig.\@ \ref{fig:Multi-volumeRefocusing}.
First, the defocus parameters are estimated from the first OCT volume by the method described in Section \ref{sec:Refocusing}.
The same phase-only spatial frequency filter [Eq.\@ (\ref{eq:2})] with these estimated defocus parameters is then applied to all 32 volumes for refocusing.

Two types of DOCT algorithms, logarithmic intensity variance (LIV) and Late OCT correlation decay speed (\OCDS), were used for DOCT imaging.
LIV is defined as the temporal variance of the dB-scaled OCT intensity signal \cite{Elsadek2020BOE}, which is mainly sensitive to the occupancy of the dynamic scatterers over all scatterers in the resolution volume \cite{Morishita2026BOE}.

\OCDS is defined as the slope of the autocorrelation decay curve of the time-sequence OCT signal within a specific delay-time range \cite{ElSadek2021BOE}.
In the present study, the delay-time range of [250.5 ms, 1504.0 ms] was used.
Before computing the slope by linear regression, the autocorrelation decay curve was kernel averaged over $2 \times 4$ pixels in the axial and lateral directions.
The \OCDS is known to be sensitive to scatterer dynamics with a specific speed that is defined by the delay-time range \cite{Morishita2026BOE}.
In our specific case, the \OCDS is sensitive to the speed around 0.11 \um/s.

\section{Experimental validation}
\subsection{Method}
\subsubsection{Samples}
Four human breast adenocarcinoma spheroids (MCF-7 cell-line) were imaged.
Each spheroid was formed by initially seeding 1,000 tumor cells.
Two spheroids were treated by an anti-cancer drug (doxorubicin; DOX), where the drug concentration was 1 \uM or 10 \uM for each spheroid.
The drug was applied on day 5 of the cultivation and the imaging was performed on day 8 (i.e., 3-day treatment time).
The other two were kept untreated and measured on day 8 and day 9.
The spheroids were cultivated with a 96-well plate but transferred to a Petri dish for the OCT imaging.

\subsubsection{OCT and DOCT imaging}
All the samples were measured by the SC-FFOCM system described in Section \ref{sec:principle}.
In addition, one non-treated spheroid was also measured by a point-scanning OCT at the 840-nm band.
For this point-scanning imaging, the lateral field was divided into eight sub-fields, and each of them was repeatedly scanned by a raster scan protocol for 32 times.
The repetition time was 202.56 ms, and the measurement time for a single sub-field was approximately 6.48 s.
The total acquisition time for the volume was 51.9 s.
Each \enface image consists of 512 (fast-scan direction) $\times$ 128 (slow-scan direction) pixels with an FOV of 0.60 mm $\times$ 0.60 mm.
Namely, the \enface pixel size (fast-scan $\times$ slow-scan directions) is 1.17 \um $\times$ 4.69 \um.
More details of the point-scanning system and the scan protocol are described elsewhere \cite{Morishita2023BOE}.

\subsection{Results}
\subsubsection{Computational refocusing}
\begin{figure}
	\centering\includegraphics[width=13cm]{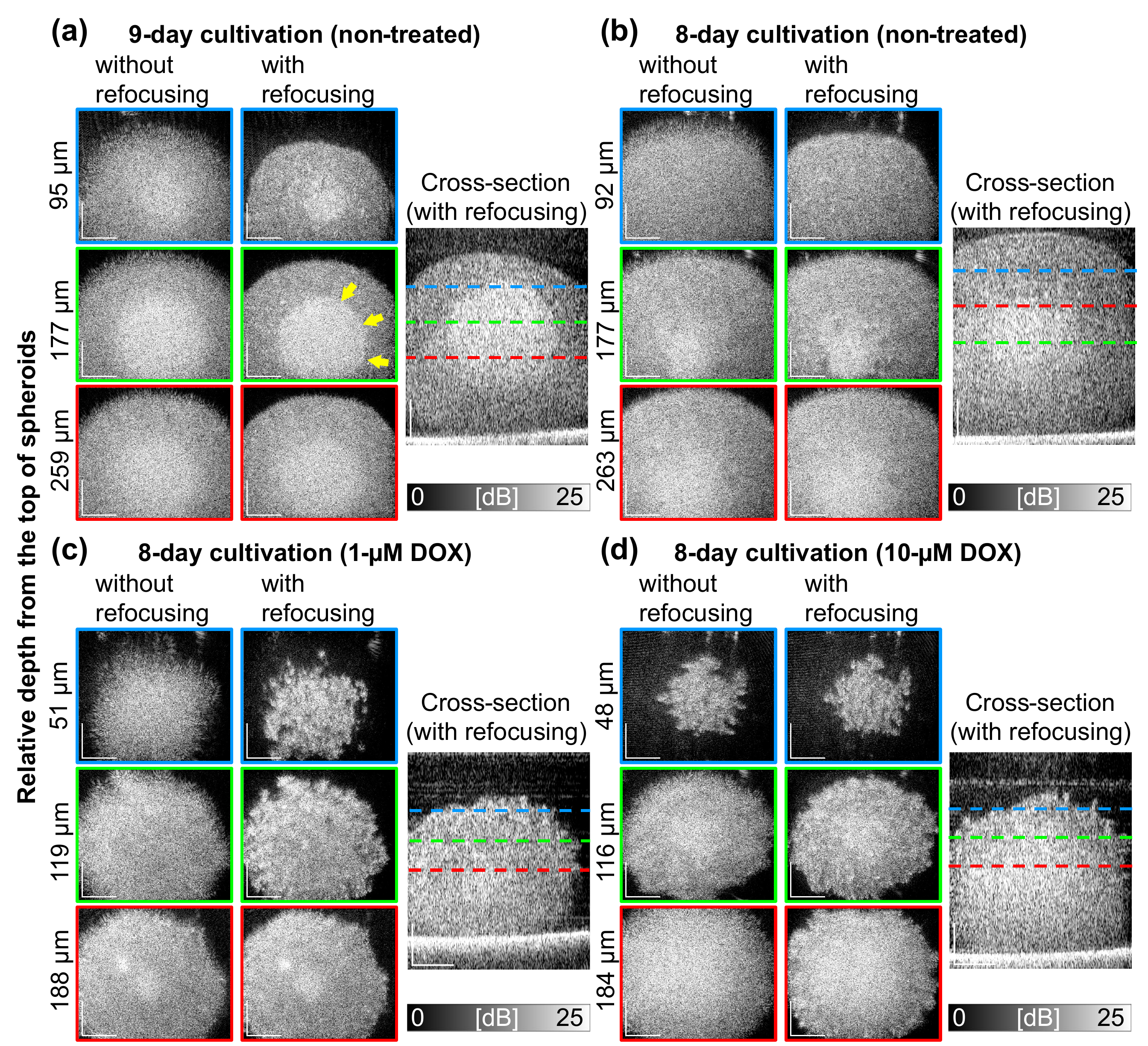}
	\caption{\textit{En face} OCT intensity images of spheroids (a)-(b) without drug administration, and (c) treated with 1 \textmu M and (d) 10 \textmu M DOX without and with refocusing and refocused cross-sectional images.
		The depths of \enface images from the top of the spheroid are described at the left side of \enface images and indicated as broken lines in cross-sectional images.
		The yellow arrowheads indicate the boundary of the core and the periphery.
		The scale bar denotes 100 \um.}
	\label{fig:Refocusing}
\end{figure}
Figure~\ref{fig:Refocusing} shows the \enface OCT intensity images without and with refocusing, and refocused cross-section images of four spheroids.
Because of the small FOV, some parts of the spheroids are out of the FOV\@.
Images without refocusing are significantly blurred and significant structures are not recognizable.
On the other hand, computational refocusing sharpens the images at all depths.
In addition, because there is no confocality in SC-FFOCM (as discussed in Section \ref{sec:confocality}), the image contrast is not significantly altered along the depth once the defocus was corrected.

After the refocusing, a clear boundary between the high-scattering core and low-scattering periphery regions is visible at 177-\um depth of the day-9 non-treated spheroid [Fig.\@ \ref{fig:Refocusing}(a)].
MCF-7 spheroids are known to form necrotic cores caused by hypoxia and nutrient deficiency \cite{COSTA20161427}.
This boundary may indicate the transition between the central necrotic core and the peripheral region.
At a deeper depth (259 \um), the core boundary becomes relatively unclear although the outer bound is clear.
This would be caused by multiple-scattering signals from the superior regions.
See Section \ref{sec:msSignal} for a more detailed discussion about the multiple-scattering effect.
In the DOX-treated samples [Figs.\@ \ref{fig:Refocusing}(c) and \ref{fig:Refocusing}(d)], the surface structures of the spheroids are broken, maybe because of the cell-membrane damage induced by DOX \cite{PilcoFerreto2016IJOnc}.

\subsubsection{DOCT imaging}
\begin{figure}
	\centering\includegraphics[width=13cm]{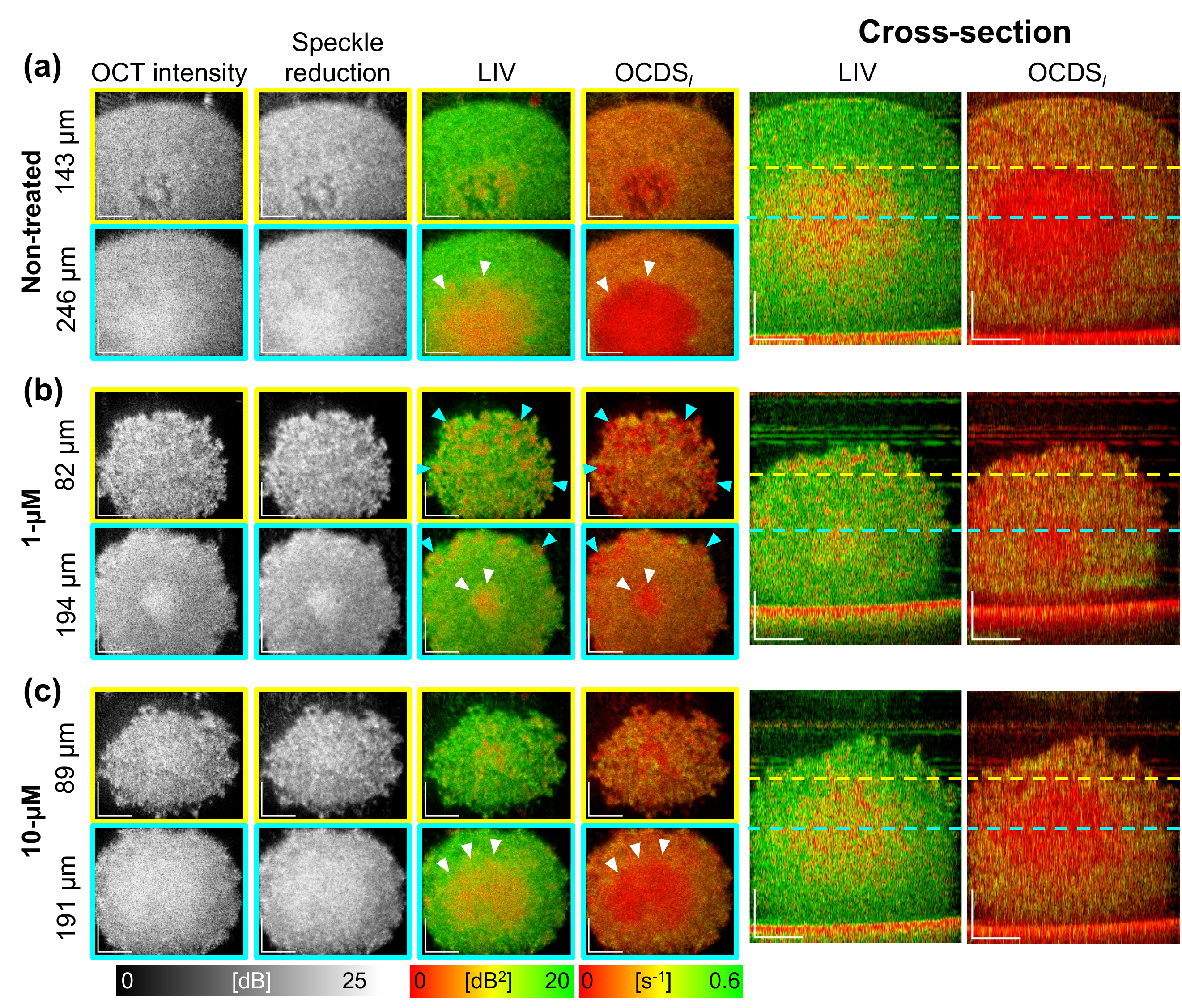}
	\caption{\textit{En face} and cross-sectional images of OCT intensity, speckle-reduced intensity, LIV, and \OCDS (a) without drug treatment, as well as with (b) 1 \textmu M and (c) 10 \textmu M DOX\@.
		The depths of \enface images from the top of the spheroid are described at the left side of \enface OCT intensity images and indicated as broken lines in cross-sectional images.
		The scale bar denotes 100 \um.}
	\label{fig:DOCT}
\end{figure}
The OCT and DOC images of three spheroids are summarized in Fig.\@ \ref{fig:DOCT}.
The second column shows the ``speckle-reduced OCT images,'' each of which was generated from a single OCT volume by the shifted-complex-conjugate-product averaging (SCCP averaging) method (details in Ref.\@ \cite{Xibo2025BiOS}).
The parameters used for the SCCP averaging method was an \enface maximum shift of $\pm$ 8 pixels (3.84 \um).
The color-coded DOCT images are the fusion of DOCT (LIV and \OCDS) and the speckle-reduced OCT image \cite{Elsadek2020BOE}, where they were used to define the hue and the brightness of the pixel, respectively.
It should be noted that the OCT intensity images used to compute the DOCT had only been refocused but are not the speckle-reduced image.

In the non-treated spheroid, both LIV and \OCDS are low in the spheroid core and high in the periphery, as shown in Fig.\@ \ref{fig:DOCT}(a).
The spheroid treated with 1-\uM DOX [Fig.\@ \ref{fig:DOCT}(b)] also exhibits the low LIV and \OCDS core (white arrowheads), but the core is smaller than that of the non-treated spheroid.
In addition, several localized low LIV and \OCDS regions are visible in the periphery (blue arrowheads).
In the spheroid treated with 10-\uM DOX [Fig.\@ \ref{fig:DOCT}(c)], a large core with low LIV and \OCDS (white arrowheads) is visible.
These results are consistent with our previous study of drug-spheroid-interaction imaging by point-scanning DOCT \cite{ElSadek2024SR}.

\subsubsection{Comparison of point-scanning OCT and SC-FFOCM}
\label{sec:Comparison_of_scanning}
\begin{figure}
	\centering\includegraphics[width=13cm]{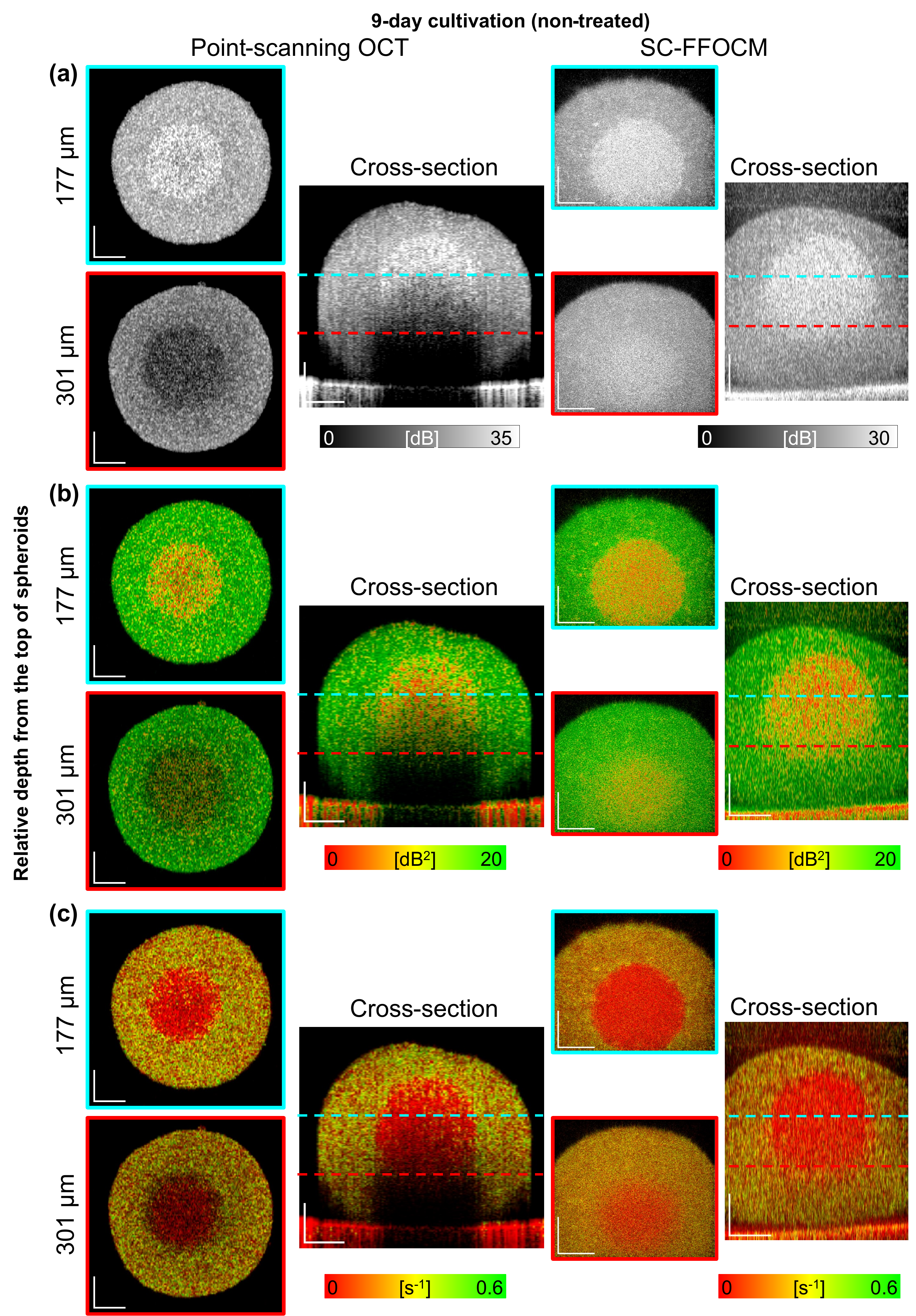}
	\caption{(a) OCT intensity, (b) LIV, and (c) \OCDS images measured in point-scanning OCT and SC-FFOCM\@.
		The depths of \enface images from the top of the spheroid are described at the left side of \enface images and indicated as broken lines in cross-sectional images.
		The scale bar denotes 100 \um.
		}
	\label{fig:dBIntandDOCTcomparison}
\end{figure}
The intensity OCT, LIV, and \OCDS images of a non-treated spheroid (day 9) were compared between point-scanning OCT and SC-FFOCM\@ [Fig.\@ \ref{fig:dBIntandDOCTcomparison}].
The intensity OCT image [Fig.\@ \ref{fig:dBIntandDOCTcomparison}(a)] is the intensity-averaged image of the 32 frames, and this image is also used as the brightness channel of the color-coded DOCT images.
The speckle-reduction method used in the previous section \cite{Xibo2025BiOS} was used because it was specifically designed for the SC-FFOCM, and its application to the point-scanning OCT is not straightforward.

In the point-scanning OCT, the region under the spheroid core became significantly dark and invisible [Fig.\@ \ref{fig:dBIntandDOCTcomparison}(a)] because of the combined effect of the signal attenuation by the strong scattering at the core and the confocal effect.
On the other hand, since the SC-FFOCM is free from the confocal effect, the region under the core is still observable.
In the DOCT images [Fig.\@ \ref{fig:dBIntandDOCTcomparison}(b)], significant differences in the overall DOCT values are not observed.
The possible effects of several different factors on DOCT contrast are discussed in Section \ref{sec:DOCTdifference}.

\section{Discussion}
\subsection{Multiple scattering}
\label{sec:msSignal}
Even after the image was sharpened by computational refocusing, the internal structure was sometimes still obscured, especially at the deep region in the tissue.
This may be because of overlaying multiple-scattering signal.
Namely, the SC-FFOCM does not have a confocal pinhole, and hence, it cannot physically reject the multiple-scattering light.

Multiple scattering can be reduced by several methods such as multi-focus averaging (MFA)\cite{zhuMultifocusAveragingMultiple2023}, wavefront modulation by using a spatial light modulator or a deformable mirror \cite{Borycki2019BOE, auksoriusCrosstalkfreeVolumetricVivo2019a, Stremplewski:19}, or spatial-coherence manipulation by multi-mode fiber \cite{Auksorius:22}.
Introduction of these methods may further improve the image quality of the SC-FFOCM.

\subsection{Tissue-induced defocus and aberrations}
In the present study, the defocus was corrected by computational refocusing.
However, the sample structure can cause additional defocus, and it varies over the lateral field.
For example, the interface between the spheroid and the surrounding medium can induce additional defocus, and it varies over the lateral position.
One possible solution is computational refocusing with small lateral sub-regions in which the defocus can be regarded as a constant \cite{auksorius_vivo_2020}.

In addition, some samples can have an irregular surface, and this induces high-order tissue-induced aberration.
These high-order aberrations are not fully correctable by the sub-region based refocusing.
High-order aberration correction or high-order phase error correction, such as adaptation of computational adaptive optics \cite{adie_computational_2012,kumar_subaperture_2013,hillmann_aberration-free_2016} might be a solution to retrieve fully diffraction-limit resolution.

Finally, our defocus model assumed the defocus amount is linear to the depth in the sample and also assumed the depth in the image is proportional to the depth in the sample.
However, if the refractive index of the sample varies along the depth, the latter assumption cannot hold.
For possible future cases in which the refractive index of the sample rigorously varies along the depth, the refocusing accuracy can be further improved by modeling the defocus amount as a high-order function of the depth.

\subsection{Future works}
The current system implementation still has some issues.
Here we discuss the possible modifications of the SC-FFOCM system as future works.

\subsubsection{Imaging depth range extension}
In the current system, the axial location of the sample is not monitored during the alignment of the sample location.
It is not easy to locate the sample within the imaging range before the acquisition.

For future applications of the system, a larger axial imaging range or a sample targeting system will be required.
By using an off-axis reference beam, the autocorrelation signal and complex conjugate mirror image can be separated from the actual signal to achieve twice the imaging depth range and remove autocorrelation artifacts\cite{hillmannOffaxisReferenceBeam2017a}.

However, the custom-built SC-FFOCM is based on a simple Michelson interferometer with a gold-mirror reference arm, modifying the reference beam is challenging.
Replacing the reference arm with a single-mode fiber would enable easier manipulation of the reference beam, such as off-axis adjustments.
Note that the phase carrier using the off-axis reference beam causes the phase sampling limit described in Section\@ \ref{sec:systemDesign} to be tightened.

\subsection{DOCT difference between point-scanning OCT and SC-FFOCM}
\label{sec:DOCTdifference}
Our experimental comparison of the point-scanning-based and SC-FFOCM-based DOCT did not show a marked difference in the DOCT contrast (Section \ref{sec:Comparison_of_scanning}).
As summarized in Table {tab:Parameters}, the point-scanning OCT and SC-FFOCM of this study share the same wavelength band, but SC-FFOCM has a significantly higher resolution.
It has been experimentally and numerically shown that the wavelength of the probe beam significantly affects the DOCT values, while the resolution has a relatively small impact \cite{fujimura_experimental_2025-1} as far as the number of scatterers in a resolution volume is not too small.
The results in the present study agree with this report.
\begin{table}[htbp]
	\caption{Parameters of the DOCT acquisition protocol. In tissue resolutions are calculated with the refractive index of the sample as 1.36.}
	\centering
	\footnotesize
	\begin{tabular}{ccc}
		& Point-scanning OCT\cite{Morishita2023BOE} & SC-FFOCM \\
		\midrule
		Center wavelength [nm]
		& 840 & 840 \\
		Lateral resolution (FWHM-intensity) [\textmu m]
		& 2.0 & 1.4 \\
		Axial resolution (FWHM-intensity) [\textmu m]
		& 3.8 (in tissue) & 4.8 (in tissue) \\
		Depth-signal acquisition time
		& 20 \us & 250 ms \\
		Repetition time
		& 202.56 ms (inter-frame) & 250.50 ms (inter-volume)\\
		Acquisition-time window for a time-sequence [s]
		& 6.28 & 7.77 \\
	\end{tabular}
	\label{tab:Parameters}
\end{table}

Although the particular comparison in the present study does not show a significant difference, there are some possible factors that may cause the difference in the DOCT values.
One factor is the number of scatterers in a resolution volume (i.e., effective number of scatterers; ENS).
If ENS becomes small, the effect of scatterer number fluctuation can be prominent \cite{Cheishvili2023OpEx}.
Under this condition, the resolution difference between the two systems can cause a difference in DOCT values.
The other possible influential factor is the ``depth-signal acquisition time.''
This is the A-scan acquisition time for the point-scanning OCT and the volume acquisition time for the SC-FFOCM, and the scatterer dynamics in the sample during this acquisition time cannot be properly captured in the OCT signal.
The depth-signal acquisition time of SC-FFOCM is significantly longer than the point-scanning OCT (20 \us versus 250 ms, see Table {tab:Parameters}).
This difference may potentially affect the DOCT values.

\section{Conclusion}
In this paper, we demonstrated SC-FFOCM imaging with a high NA that is incorporated with computational refocusing, and DOCT imaging.
This imaging modality was applied for imaging of cancer spheroid and revealed fine structural and dynamic properties of the sample over a long depth region.
In addition, the comparison study with a point-scanning OCT demonstrates better image penetration of SC-FFOCM\@.
These findings suggest that SC-FFOCT can be a useful tool for structural and functional imaging of thick \invitro samples.

\begin{backmatter}

\bmsection{Funding}
Core Research for Evolutional Science and Technology (JPMJCR2105);
Japan Society for the Promotion of Science (21H01836, 22F22355, 22KF0058, 22K04962, 23KF0186, 24KJ0510);
Chinese Scholarship Council (202106845011);
National Natural Science Foundation of China (62005123);
Natural Science Foundation of Jiangsu Province (BK20190455).

\bmsection{Acknowledgment}
Makita corresponds to the computational refocusing methods, while Yasuno corresponds to the other parts of the manuscript.
Tateno and Zhu equally contributed to this manuscript.

\bmsection{Disclosures}
Tateno, Zhu, Komeda, Ishikawa, Wang, El-Sadek, Morishita, Makita, Yasuno: Sky Technology(F), Nikon(F), Kao Corp.(F), Topcon(F), Panasonic(F), Santec (F), Nidek (F);
Furukawa, Matsusaka: None.
Tateno is currently employed by Kowa Co., Ltd..
	
\bmsection{Data availability} Data underlying the results presented in this paper are not publicly available at this time but may be obtained from the authors upon reasonable request.

\end{backmatter}

\bibliography{SC-FFOCM}

\end{document}